\pdfoutput=1
\documentclass[sigconf,nonacm]{acmart}

\AtBeginDocument{%
  }

\usepackage{algorithm}
\usepackage{subcaption}
\usepackage{array}
\usepackage{threeparttable}
\usepackage{adjustbox}
\usepackage{algpseudocode}
\usepackage[most]{tcolorbox}
\usepackage{cleveref}
\usepackage{enumitem}

\newcommand{\eg}{\textit{e.g.,\ }}
\newcommand{\ie}{\textit{i.e.,\ }}

\tcbset{enhanced, boxrule=0pt, frame hidden, borderline west={4pt}{0mm}{white!75!black}, colback={black!5!white}}

\begin{document}

\title{Local Inference of LLMs on Long Contexts:\\ A Data Systems Perspective}

\author{Mattias D{\"u}rrmeier}
\orcid{0000-0000-0000-0000}
\affiliation{%
  \institution{University of Fribourg}
  \streetaddress{Bd de Pérolles 90}
  \city{Fribourg}
  \country{Switzerland}
}
\email{mattias.duerrmeier@unifr.ch}

\author{Alberto Lerner}
\orcid{0000-0000-0000-0000}
\affiliation{%
  \institution{Computing Flows GmbH}
  \city{Fribourg}
  \country{Switzerland}
}
\email{lernera@computingflows.com}

\renewcommand{\shortauthors}{D{\"u}rrmeier \& Lerner}
\renewcommand{\shorttitle}{Local Inference of LLMs on Long Contexts: A Data Systems Perspective}

\begin{abstract}
Large language models can solve complex problems across a variety of subjects, such as mathematics and coding. 
For such complex analytical tasks, models with a long context perform better. 
Serving models with long context is currently mostly done on powerful systems with multiple GPUs, which are too expensive for most users. 
An alternative is local inference, or serving a model on a user's own machine, which removes per-token costs and rate limits, besides keeping sensitive data private and giving users full control over the models they run. 
The common understanding, however, is that local LLMs need compression (\eg quantization) to run on a single GPU. 

In this paper, we set out to study the limitations that prevent serving long-context models for single users, besides the capacity issue. 
Rather than characterizing a single model on one GPU, we analyze how different models behave across several GPU generations. 
This lets us uncover not just individual bottlenecks but broader trends in where inefficiencies arise. 
Our analysis confirms that, as expected, self-attention dominates the runtime for long-context inference. 
More unexpectedly, our evaluation reveals that the current structure of self-attention leaves key GPU resources significantly underutilized, and the newest GPU is often the least well served. 
Many of these findings are fundamentally data management problems, from paging the KV cache to improving cache utilization and synchronization in the GPU, which is why we bring this study to the Data Systems community. 
To engage this community, we present our findings as a series of experimental observations and formulate recommendations for future work.
\end{abstract}

\maketitle

\section{Introduction}
\label{sec:introduction}

Transformers~\cite{vaswani2017attention} are currently the leading model architecture for a wide range of tasks.
This phenomenon is underscored by a study from Jouppi et al., the creators of Google’s TPU, that analyzed the relative prevalence of different model architectures~\cite{jouppi2026google}.
In 2026, transformers accounted for 76\% of the models deployed on TPUs, with their share rising steadily since the hardware's introduction.
By contrast, all other model architectures saw lower utilization.
The second most common, multi-layer perceptrons, comprised only 11\%.
Much of this success is driven by large language models (LLMs), the transformer's most consequential application~\cite{radford2018improving, radford2019language}.
Trained on large corpora, they can solve diverse tasks such as logic and math problems~\cite{ke2025survey, liu2024mathbench}, write and run code~\cite{jiang2026survey, jimenez2024swe, jain2025livecodebench}, or evaluate research ideas and write papers~\cite{lu2026endtoend}, to name a few.

Many of these tasks can now be performed with open-weight models, allowing users to run inference on their own hardware~\cite{grattafiori2024llama3, deepseek2026deepseekv4, openai2025gptoss, qwen2025qwen3}.
These models, however, have a very high resource consumption.
The majority of this demand comes during the decode phase of the LLM, where the conversation is generated one token at a time, each token requiring a full pass through the model (\Cref{sec:background}).
It is ultimately the ability to generate large numbers of tokens through the decode phase that allows more complex analytical tasks to be performed.
Similarly to the scaling observed during training~\cite{kaplan2020scaling}, inference scaling shows that LLMs perform better when they spend more time reasoning about the problem~\cite{ke2025survey}.

Despite this, the performance profile of the single-user (\ie locally run) decode phase has received little attention.
Prior characterization work has largely targeted high-throughput, multi-user serving~\cite{recasens2024pareto, recasens2025mind}, where the aim is to maximize the number of queries processed at once.
In that regime, inference is known to be memory-bound~\cite{recasens2025mind, gholami2024ai}.
In the single-user setting, memory capacity becomes the main limitation because the model's state might not fit on a single GPU. 
The typical solution is to either compress the state through quantization~\cite{dettmers2022llm, frantar2022gptq,hooper2024kvquant} or to page parts of the state in and out of the GPU~\cite{sheng2023flexgen, jiang2025neo}.
Doing so, however, prevents the remaining causes of performance loss from being properly understood.

In this paper, we address this gap by providing a performance characterization study of LLMs under a single-user workload.
Our central objective is to locate bottlenecks as models and hardware evolve.
To achieve this, we examine how performance changes both with different models and model sizes, and as we move across three successive generations of GPUs (\Cref{sec:methodology}). 
The measurements we make are at the level of the GPU kernels produced by the most common deep learning library and its compiler, PyTorch~\cite{ansel2024pytorch2}.
This allows us to introspect not only how each kernel performs, but also whether it reflects an efficient compilation of the model. 
What we found are many inefficiencies that, if addressed, can unlock further performance of local LLMs.

Our performance characterization study comprises two parts.
The observations come first, and are based on how the decode phase behaves across different models and GPUs (\Cref{sec:gpt-models} and \Cref{sec:qwen3-models}).
It finds that, for instance, the newest GPU is often the least well utilized, as its attention kernel fails to translate greater hardware capability into faster decoding.
Our study's recommendations follow, addressing the limitations these observations expose.
For instance, there are several opportunities to better utilize the GPU's resources under the single-user regime  (\Cref{sec:recommendations}).

Taken together, the observations and the recommendations put three larger questions within reach (\Cref{sec:discussion}).
The first is whether the inefficiencies we found are ones this community has met before.
We believe they are, as evidenced by the parallels found in this community’s prior work on GPU query processing.
The second is whether the compression techniques we mentioned above remain necessary once the device is used well.
We argue they may not have as important a role as they currently do.
As we will see, paging techniques we discuss here may offer a way to fit a model on the device without the loss in downstream performance that compression brings~\cite{jaiswal2024compressing}.
The third asks what kind of system would be required to implement these recommendations.
We argue that developing a system specifically tailored to this workload is more promising than adapting these techniques to systems designed for different purposes.

In summary, this paper makes the following contributions:

\begin{itemize}[leftmargin=*]
  \item We profile a single decoder block at the level of its compiled kernels, which makes models larger than the device under study measurable, and apply this across two model families, three GPU generations, and context lengths reaching one million tokens.
  \item We find the reasons why no device is overwhelmed by this workload, and that all three have resources to spare, and we establish where that spare capacity lies, in which kernels, on which hardware, and at what context lengths.
  \item We derive from these findings four recommendations, which together describe the axes along which a system specialized for this workload could be built.
  \item We relate each of those axes to prior work in which this community addressed the same difficulty in GPU query processing.
\end{itemize}

\section{Background}
\label{sec:background}

A large language model generates text one token at a time, repeatedly evaluating the same stack of decoder blocks to produce each token.
To run it, the block is specified as a sequence of mathematical operations and subsequently compiled into execution units known as kernels, the unit of code execution on GPUs.
This section describes both the operations that define the decoder block and the kernels into which they are compiled.

\subsection{The Model Definition}
\label{ssec:model-definition}

\begin{figure}[h]
  \centering
  \includegraphics[width=\linewidth]{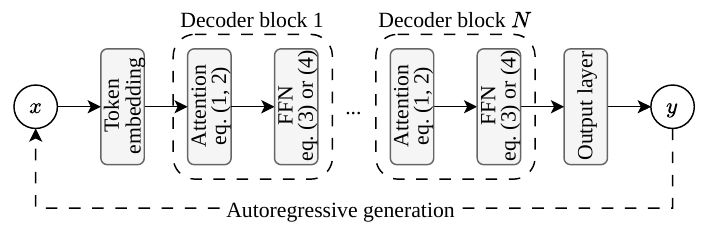}
  \caption{Overview of the autoregressive generation of a large language model.}
  \label{fig:llm-overview}
\end{figure}

A large language model is a decoder-only transformer model composed of a sequence of $N$ decoder blocks, each containing an attention layer and a feedforward network (FFN). The model organizes these components as shown in Figure~\ref{fig:llm-overview}.

The attention layer first projects its input into three matrices: the queries, keys, and values.
It then computes a similarity between tokens through two dot products, with a softmax applied in between.
\begin{equation}\label{eq:attn}
    \begin{split}
    Q, K, V &= x W_{Q K V} \\
 \text{Attention}(Q, K, V) &= \text{softmax}(\frac{Q K^T}{\sqrt{d_k}}) V
    \end{split}
\end{equation}
Multi-head attention (MHA) performs several such computations in parallel, each head attending to a different part of the sequence.
\begin{equation}\label{eq:mha}
	\begin{split}
		\text{MultiHeadAttention}(Q, K, V) &=
		\text{concat}(\text{H}_1, \dots, \text{H}_h) W_{out} \\
		\text{where}\ \text{H}_i &= \text{Attention}(Q_i, K_i, V_i)
	\end{split}
\end{equation}
The FFN that follows applies two linear projections with a non-linear activation function between them, such as GELU~\cite{hendrycks2023gelu}.
\begin{equation}\label{eq:ffn}
 \text{FFN}(x) = \text{act}(x W_{1} + b_1) W_{2} + b_2
\end{equation}
The vast majority of a model's parameters reside in these two layers, and so does most of the work of inference.
Finally, a normalization layer is added before each of the two layers~\cite{ba2016layer, zang2019root} and a residual connection around each of them~\cite{he2016deep}.

Modern LLM architectures sometimes use mixture of experts (MoE), a sparse version of the FFN layer~\cite{shazeer2017outrageously, fedus2022switch, jiang2024mixtral}.
MoE uses conditional computations to activate only a subset of the network for each input.
It is composed of $n$ experts, each a FFN, and a gate $G$.
The gating function evaluates the $k$ most relevant experts for the input.
The gate is learned through a linear projection.
\begin{equation}\label{eq:moe}
    \begin{split}
    \text{MoE}(x) &= \sum_{i}^{n} G(x)_i E_i(x) \\
    G(x) &= \text{softmax}(\text{TopK}(x W_{g}), k)
    \end{split}
\end{equation}
where $\text{TopK}(v, k)_i = v_i$ if $v_i$ is among the $k$ selected experts, and $- \infty$ otherwise.
The coefficients weight the contribution of the selected experts, and the softmax layers ensure that the sum of them equals one.
This allows packing more weights in a model but keeps the computational cost lower compared with a dense approach.

Inference proceeds in two phases~\cite{li2024llm, zhou2024survey}.
Prefill processes the full input prompt and produces the first token, while decode generates each subsequent token from the ones before it.
The two phases behave differently on the hardware, and this paper studies decode, which dominates the cost of the long generations produced by reasoning and agentic workloads.
Decode is made efficient by a key-value (KV) cache.
Rather than recomputing the keys and values of all previous tokens at every step, the model stores them and computes only the query, key, and value of the current token, as shown in Algorithm~\ref{algo:background:llm-inference}.

\begin{algorithm}[h]
 \caption{LLM Inference with KV Cache}
 \label{algo:background:llm-inference}
 \begin{algorithmic}[1]
  \Procedure{generate}{model, $[x_1, ..., x_n]$, max length $M$}
   \State logits, $K, V$ := forward(model, $[x_1, ..., x_n]$) \Comment{prefill}
   \State $y_{i}$ := selectNextToken(logits) \Comment{greedy, top-k, etc.}
   \State $y := [y_{i}]$

   \While{$y_i \neq \text{EOS}$ \textbf{and} $\text{len}(y) < M$}
    \State logits, $(k_{i}, v_{i})$ := forward(model, $y_i, K,V$) \Comment{decode}

    \State $K$ = append($K, k_{i}$) \Comment{update cache}
    \State $V$ = append($V, v_{i}$)

    \State $y_{i+1}$ = selectNextToken(logits)
    \State $y$ = append($y, y_{i+1}$)
    \State $y_i = y_{i+1}$
   \EndWhile
   \State \textbf{return} $y$
  \EndProcedure
 \end{algorithmic}
\end{algorithm}

As a result, the cost of a decode step stays constant across most of the block, and grows only in the attention layer, where the current token must attend to a KV cache that lengthens with every token generated.
This asymmetry, a fixed cost everywhere except an attention cost that grows with context, plays a central role in the analysis that follows.

The two layers of the decoder block have since been revised to raise capacity while holding computation down, and these revisions add many smaller operations around the two main ones.
Grouped-query attention (GQA)~\cite{ainslie2023gqa} lets several query heads share a single key-value head, which shrinks the KV cache and the memory it consumes during decode.
SwiGLU~\cite{shazeer2020glu} modifies the two-layer FFN by changing GELU to SiLU~\cite{elfwing2018sigmoid, ramachandran2018searching} and adding an extra linear projection.
Attention can also be augmented with rotary position embeddings (RoPE)~\cite{su2024roformer}, which encode token positions by rotating the queries and keys before the dot product.
Each of these changes improves the model, yet each also introduces additional small operations, a point we return to when we examine how a decoder block executes on the GPU.

\begin{figure*}[h]
	\centering
	\includegraphics[width=0.85\textwidth]{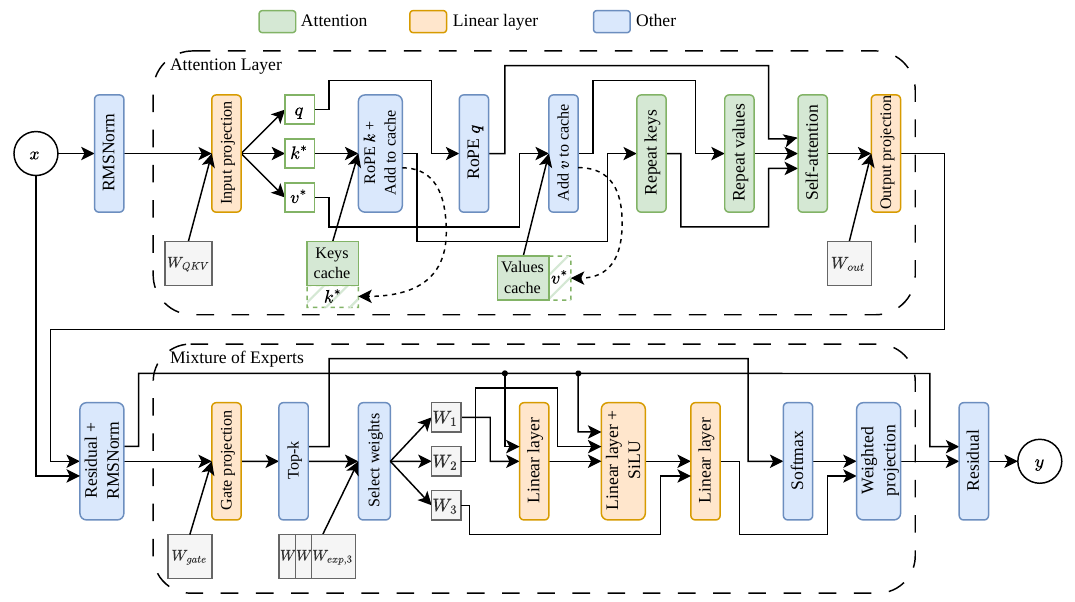}
	\caption{Per kernel view of the decoder block in GPT-OSS, compiled for the RTX PRO 6000 Blackwell.}
	\label{fig:llm-compiled-view}
\end{figure*}

\subsection{The Model Execution}
\label{ssec:model-execution}

A GPU is essentially a parallel device made of a large number of symmetric multiprocessors, each of which contains different types of processing cores~\cite{hwu2022programming}.
Two types of core matter for our workloads.
First, the CUDA cores, which perform general arithmetic, such as the softmax in attention and the activation in the FFN.
Second, the Tensor cores, which are specialized units that perform the matrix multiply and accumulate operations (MMA) at the heart of LLMs, including those in Equation~\eqref{eq:attn} and Equation~\eqref{eq:ffn}.

The decision on how to translate a model's definition into code rests on the chosen framework's compilation stack~\cite{frostig2018compiling, ansel2024pytorch2}.
This process lowers a block definition into a graph of tensor operations.
These include matrix multiplications, elementwise functions, and reductions, which together realize the high-level structure of the block.
Each element of the graph could technically become a kernel, a small program that the GPU runs, launched to perform one operation across many threads at once.
The GPU executes these kernels largely one after another, so a block runs as a sequence of them, though independent kernels can sometimes overlap.

The compilation process, however, optimizes the graph before mapping it to kernels.
It merges certain element-wise operations, such as an activation or a scaling, into the neighboring matrix multiply.
It also searches for subgraphs it knows how to represent as a single kernel.
A notorious example is FlashAttention, which fuses attention's two matrix multiplications and the softmax between them into a single kernel~\cite{dao2023flashattention2}.
Each such merge removes an intermediate result that would otherwise be written to memory and read back by the next kernel.
Many operations, however, are left on their own, either because no safe merge exists or because no pattern matches them.
A block therefore compiles into a long sequence of kernels of uneven size.

To give a concrete example, the attention and FFN operations of Equations~1, 2, and 4 compile into the kernel structure shown in \Cref{fig:llm-compiled-view}.
The block compiles into several kernels, which are very diverse in terms of the resources each consumes, their sizes, and runtimes. 
The figure also shows that some kernels take external inputs, such as the weight matrices and the KV cache.
The cache is updated within the attention layer, where the new key and value are appended to the cache.

One important aspect for our purposes is the memory cost of connecting these kernels.
The output of one kernel is written to global memory and read back by the next, so a long chain of kernels means repeated round trips through the memory hierarchy.
Each of these transfers consumes memory bandwidth, one of the two resources, alongside compute, whose use we measure throughout this work.

\section{Methodology}
\label{sec:methodology}

Characterizing LLM inference for local usage presents several challenges.
First, these models are constantly evolving. 
Focusing only on the latest model provides a snapshot but offers no assurance that our findings will apply to future versions.
Second, while different models may share the general LLM structure, they often differ in the details of the attention block and feedforward network, which are the most critical components.
From the outset, we recognized that a proper characterization of local LLM execution requires studying long-term trends instead of isolated observations.
Therefore, our experimental methodology emphasizes broader patterns across models and hardware, rather than focusing on an individual model and GPU combination.

Studying these trends across hardware requires running the same models on different GPUs, yet the largest models may not fit, especially on older devices.
A model, however, is a stack of identical decoder blocks. 
We therefore implement and profile a single block, which behaves the same as any other in the model. 
This removes the size barrier, letting us run any models on any GPU.
It also lets us write every block in the same framework, rather than compare models across the frameworks/libraries they ship in.

We study two families of models, chosen so that each isolates a different source of change.
The first is the GPT family, spanning GPT-2~\cite{radford2019language}, GPT-3~\cite{brown2020language}, and GPT-OSS~\cite{openai2025gptoss}.
Across these three, the decoder block itself has changed, from multi-head attention and a dense feedforward network to grouped-query attention, mixture of experts, and added components such as rotary position embeddings.
This family lets us see how architectural model evolution has reshaped the workload.
The second is the Qwen3 family~\cite{qwen2025qwen3}, in its 4B, 8B, and 32B variants.
These share one modern architecture and differ only in their dimensions, which isolates the effect of model size.
Underneath both model families, we vary the hardware, profiling on three generations of NVIDIA GPUs whose specifications appear in \Cref{tab:gpu-specs}.

\begin{table}[h]
	\caption{Specifications for the three GPUs evaluated~\cite{nvidiacorporation2018nvidia, nvidiacorporation2023nvidia, nvidiacorporation2025nvidia}.}
    \label{tab:gpu-specs}
    \centering
    \begin{adjustbox}{max width=\linewidth}
        \begin{tabular}{l m{5em} m{6em} m{6em}}
            \toprule
			Graphics Card           & Quadro\newline RTX 6000 & RTX 6000 Ada Generation & RTX PRO 6000 Blackwell\\
            \midrule
            Microarchitecture       & Turing            & Ada Lovelace              & Blackwell\\
            \#CUDA Cores            & 4,608             & 18,176                    & 24,064\\
            \#Tensor Cores          & 576               & 568                       & 752\\
			L2 cache (MB)			& 6					& 96						& 128\\
            Memory (GB)             & 24 GDDR6          & 48 GDDR6                  & 96 GDDR7\\
            Memory Interface        & 384-bit           & 384-bit                   & 512-bit\\
            Mem Bandwidth (GB/s)    & 672               & 960                       & 1,792\\
            PCIe Interface          & 3.0 x16           & 4.0 x16                   & 5.0 x16\\
            TPD                     & 260 W             & 300 W                     & 600 W\\
            \bottomrule
        \end{tabular}
    \end{adjustbox}
\end{table}

The two families are paired with the hardware differently.
Each GPT model is profiled on the GPU of its era: GPT-2 on Turing, GPT-3 on Ada, and GPT-OSS on Blackwell, reproducing how model and hardware advanced together.
The Qwen3 models, in contrast, are run on all three GPUs, which holds the architecture fixed and exposes the effect of hardware alone.

Across every pairing, we profile the decode phase one kernel at a time with Nsight Compute~\cite{ncu}.
For each kernel, we record its runtime, its compute, and its memory bandwidth.
We also examine the kernels the compiler produced and whether they are the result of kernel fusion.
Two conditions keep these measurements comparable:
we profile from a cold cache, so no kernel benefits from data another left behind;
and we leave the GPU clocks unlocked, matching how these devices behave in normal use.

\begin{figure*}[t]
	\centering
	\includegraphics[width=\linewidth]{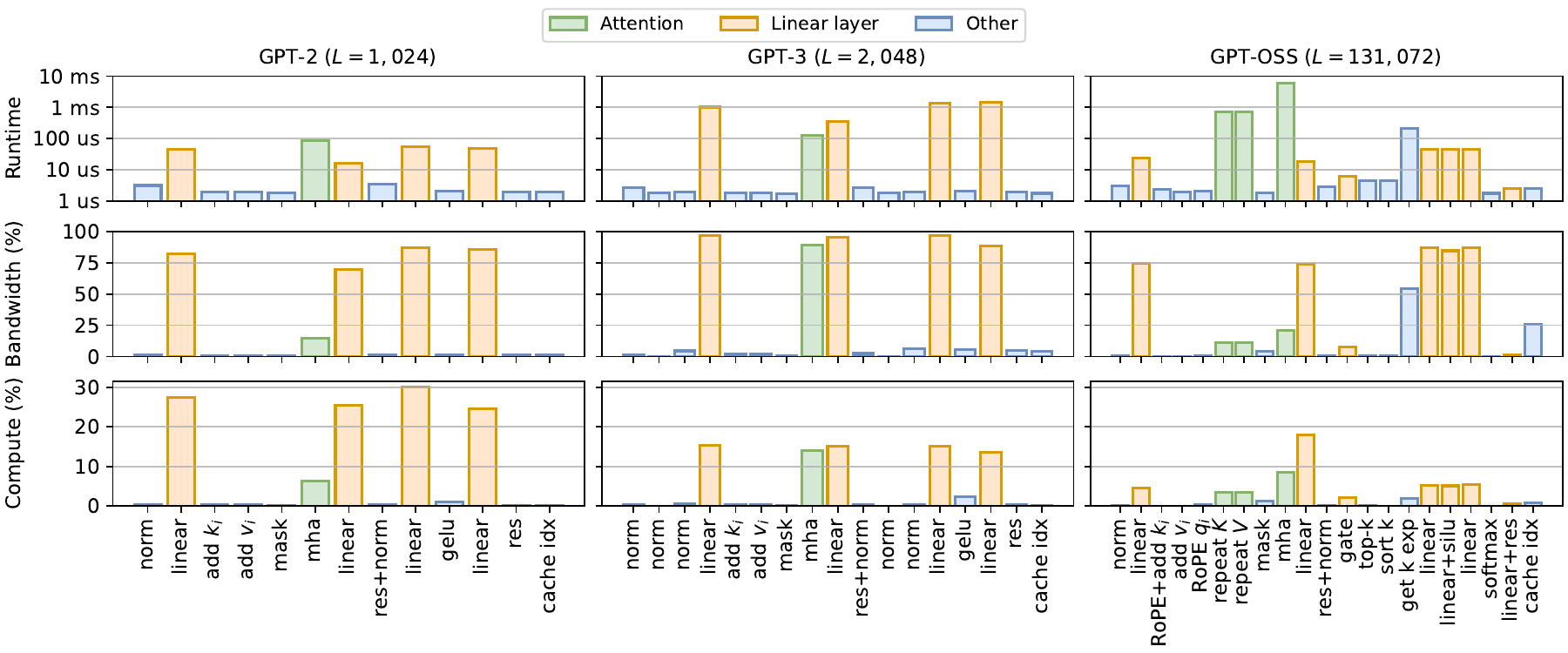}
	\caption{Runtime, bandwidth, and compute used for each GPT model at its maximum context length.}
	\label{fig:gpt-models:max-seq-len}
\end{figure*}

\section{GPT Models Evolution}
\label{sec:gpt-models}

For our first study, we analyze the three decoder blocks of the GPT family of models, shown in Table~\ref{tab:gpt-models}.
We profile each model on a different GPU: GPT-2 on Turing, GPT-3 on Ada Lovelace, and GPT-OSS on Blackwell, the leading GPU generation available at the time of its launch.
This configuration resembles the inference scenario used at the release of each model.

\begin{table}[h]
    \centering
    \begin{threeparttable}
    \caption{Block configuration of the GPT models.}
    \label{tab:gpt-models}
    \begin{tabular}{lcccccc}
        \toprule
		Model       						& $d_{model}$   & $n_{heads}$        & $d_{head}$ 	& $d_{ff}$			& Context\\ %
        \midrule
		GPT-2~\cite{radford2019language}	& 1,600         & 25                 & 64        	& 6,400  			& 1,024\\
		GPT-3~\cite{brown2020language}		& 12,288        & 96    		     & 128 			& 49,152			& 2,048\\
		GPT-OSS 120B~\cite{openai2025gptoss}& 2,880         & 64/8\tnote{$\ast$} & 64			& 2,880\tnote{$\dagger$} 	& 131,072\\
        \bottomrule
    \end{tabular}
    \begin{tablenotes}
        \item [$\ast$] Uses GQA: \#query heads/\#key-value heads
        \item [$\dagger$] Uses MoE: 128 experts, top-$k=4$; each with $d_{ff} = d_{model}$
    \end{tablenotes}
    \end{threeparttable}
    \vspace{-1em}
\end{table}

\subsection{Kernel Analysis}
\label{ssec:gpt:complexity}

\Cref{fig:gpt-models:max-seq-len} reports the runtime, bandwidth, and compute of every kernel in a decoder block, and each model uses the maximum context length set at training.
For now, we discuss only the horizontal axis of the chart, where each tick represents one kernel.
We will return to the bars and what they convey in \Cref{ssec:gpt:runtime}.
We observe that the kernel count grows with each generation, from thirteen kernels in GPT-2 to seventeen in GPT-3 and twenty-one in GPT-OSS.
GPT-2 and GPT-3 share one architecture, so the growth there comes from scale alone, as GPT-3's normalization splits into three kernels each time it appears, a process known as fission~\cite{hijma2023optimization}.

GPT-OSS is where new structure is introduced, as seen in \Cref{fig:llm-compiled-view}.
Grouped-query attention, mixture of experts, and a modern positional encoding technique each add operations absent from the older models, among them applying RoPE, the gate, the top-$k$, and the sort.
The compiler fuses some of these operations into single kernels, yet many remain kernels of their own, therefore making their count climb.
This reflects a tendency of making models perform a more elaborate set of operations at each iteration.
This will have implications, as we will see, in whether the compiler can generate a sensible set of kernels to represent the model.

\begin{tcolorbox}
	\textbf{Observation \#1:}
    Modern LLM architecture, like the one used by GPT-OSS, results in a substantial increase in the number of kernels.
    Few of them are fused by the compiler.
\end{tcolorbox}

\subsection{Runtime Analysis}
\label{ssec:gpt:runtime}

We now turn to the bars of \Cref{fig:gpt-models:max-seq-len}, which report the runtime, bandwidth, and compute of each kernel.
The kernels fall into the same three groups that \Cref{fig:llm-compiled-view} already identified: the linear projections, the attention computation, and the remaining operations of the block.
Reading the bars by group makes the shape of each block's cost visible.

In GPT-2 and GPT-3, that cost concentrates almost entirely in the linear projections.
Their kernels carry the longest runtimes and consume most of the available bandwidth, reaching 70\% to 80\% in GPT-2 and near the device peak in GPT-3, while the remaining operations use little of the device.
Attention is present but secondary, a single kernel among the taller projections, with its bandwidth use near 15\%.
This holds only for the context lengths measured here, but changes as the context grows.
We note that despite GPT-3 being 7.68$\times$ larger than GPT-2, it uses only about half as much compute.
The two models are bandwidth-bound in their projections, so what sets their compute utilization is how much compute each device carries for the bandwidth it offers.
The newer GPU carries far more, so the same class of kernel occupies a smaller share of it.

In GPT-OSS, the performance profile becomes more diverse because of the increase in the number of kernels.
The preparation and routing steps, including the repeat operations that feed attention and the gate of mixture of experts, have shifted from making a negligible contribution to accounting for a now much more substantial portion of both runtime and bandwidth.
The linear projections remain prominent, although they now account for a smaller share of compute, thanks to the increase in GPU resources for the latest generation. 

\begin{tcolorbox}
    \textbf{Observation \#2:} 
    In older models, the runtime and bandwidth of a block concentrate almost entirely on its matrix multiplications. 
    Modern architectures surround these with preparation and routing kernels, such as the repeat steps that feed attention and the gate of mixture of experts. 
    The matrix multiplications still dominate, but their share of the profile is no longer overwhelming.
\end{tcolorbox}

\subsection{Roofline Analysis of Attention}
\label{ssec:gpt:rooflines}

Attention is the one kernel whose cost grows with the context, since each new token must attend to all the tokens before it.
We therefore measure it in isolation from the rest of the block, across context lengths increasing in powers of two up to one million tokens.
This range extends beyond what any of the three models supports, but the measurement remains faithful, as the kernel performs the same operations and moves the same data at a given context length whether or not the model was built to run there.
We assess the result with the roofline model~\cite{williams2009roofline}, which bounds a kernel by the memory bandwidth and the peak compute of the device.

\begin{figure}[h!]
	\centering
	\includegraphics[width=\linewidth]{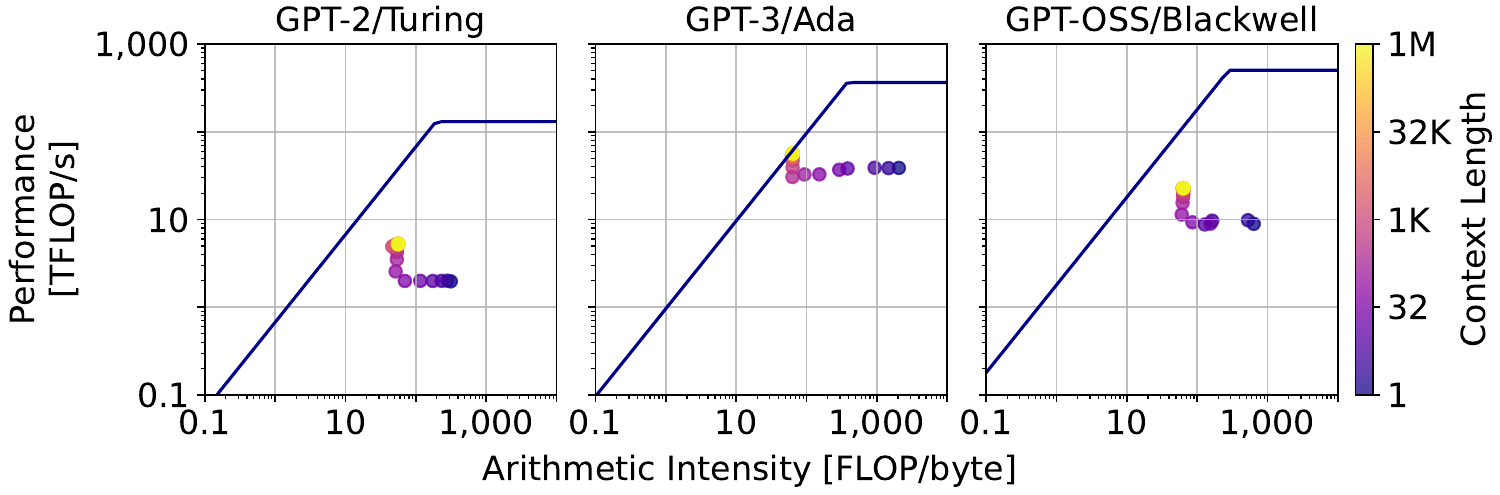}
	\caption{Attention kernel rooflines for the GPT models.}
	\label{fig:gpt-models:attention-rooflines}
    \vspace{1.5em}
\end{figure}

The three panels of \Cref{fig:gpt-models:attention-rooflines} reveal how the attention kernel behaves as the context grows.
At first, the points shift left as arithmetic intensity declines.
This occurs because the amount of data read from the KV cache increases with each token, while the computational work performed per byte remains constant.
This trend continues until the cache becomes large enough to provide sufficient parallel work for the device.
Beyond this point, the curve bends upward, which we attribute to the additional operations raising the kernel's throughput.

This increase in throughput, however, does not bring the kernel to either limit of the device.
For GPT-2 and GPT-OSS, the measurements remain well below both the bandwidth and the compute bounds at every context length, including one million tokens.
GPT-3 is the exception, approaching the bandwidth bound at its longest contexts.
The difference follows from the attention configuration, as GPT-3 uses more attention heads and larger ones, which raises the work available to the device, while the smaller heads of the other two models leave it underused.
Because bandwidth remains available and the kernel is not compute-bound, the attention computation could in principle be made faster, an opportunity we return to in \Cref{sec:recommendations}.

\begin{tcolorbox}
    \textbf{Observation \#3:} 
    Across all context lengths, the attention kernel does not reach either the bandwidth or compute limits of the device, except in the case of GPT-3. 
    Even at one million tokens, available bandwidth is not fully utilized. 
    This suggests that further improvements should focus on the kernel itself rather than the hardware.
\end{tcolorbox}

\subsection{Attention Runtime Crossovers}

\begin{figure}[h!]
	\centering
	\includegraphics[width=0.9\linewidth]{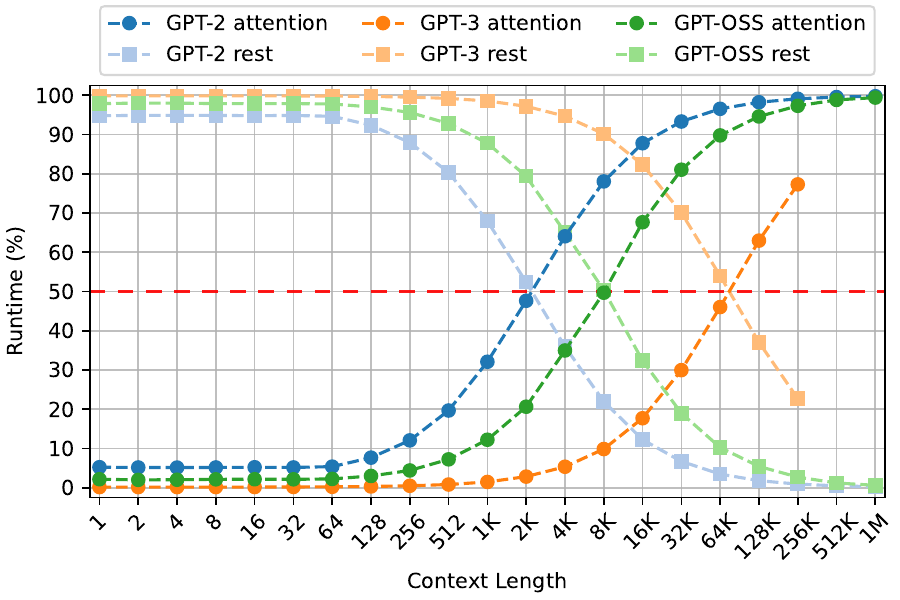}
	\caption{Runtime comparison between self-attention and the rest of the block for the GPT models.}
	\label{fig:gpt-models:attention-crossovers}
\end{figure}

As the context length grows, self-attention must be computed between the current token and a larger number of previous ones.
To study how the runtime of attention varies compared with the rest of the decoder block, we profile each model and device pair for context lengths from one to one million tokens.
Figure~\ref{fig:gpt-models:attention-crossovers} shows the contribution in runtime of self-attention relative to the rest of the model for increasing contexts.

The runtime of the three models follows an S-curve as the size of the context grows.
When working with hundreds of thousands or one million tokens, attention accounts for more than 90\% of the runtime on a modern model such as GPT-OSS.
This finding emphasizes the need for an optimized attention kernel for long contexts, as attention constitutes the vast majority of the runtime.

GPT-2 reaches the crossover at 2,048 tokens, just after its maximum context length.
As this model has the smallest dimensions, self-attention rapidly dominates the runtime.
In contrast, GPT-3 reaches this point at a much later stage.
For the largest context measured of 256K tokens before running out of device memory, attention accounts for almost 80\% of the runtime.
The larger dimensions used by this LLM incur a higher runtime for the rest of the model.
In particular, the runtime is much longer for the linear projections of this model, as seen in Figure~\ref{fig:gpt-models:max-seq-len}.

For a modern architecture, the attention computation of GPT-OSS accounts for 50\% of the execution time at a context of only 8K tokens.
This highlights the benefits of modern architectures, with results closer to those of GPT-2.
While the sparsity of MoE decreases the runtime of the rest of the block, the preparation of the KV cache increases the runtime of the attention computation.

\begin{tcolorbox}
	\textbf{Observation \#4:}
    For very long contexts, self-attention consistently dominates the execution time, regardless of model or device.
	Because GPT-2 uses a smaller architecture and GPT-OSS a more modern one, attention accounts for more than 50\% of the runtime at earlier context lengths.
\end{tcolorbox}

\section{Modern Language Models}
\label{sec:qwen3-models}

The previous section followed a family of architectures as it changed across generations, pairing each model with the hardware of its time.
This section holds the architecture fixed and varies the hardware instead.
We study the Qwen3 family, three models that share one modern design and differ only in size, shown in \Cref{tab:qwen3-models}.

\begin{table}[h]
	\caption{Block configurations of the Qwen3 models.}
	\label{tab:qwen3-models}
	\centering
	\begin{tabular}{lcccc}
		\toprule
		Model		& $d_{model}$	& $n_{heads}$ (Q/KV)	& $d_{head}$ 	& $d_{ff}$\\
		\midrule
		Qwen3-4B 	& 2,560			& 32/8					& 128			& 9,728\\
		Qwen3-8B 	& 4,096			& 32/8					& 128			& 12,288\\
		Qwen3-32B 	& 5,120			& 64/8					& 128 			& 25,600\\
		\bottomrule
	\end{tabular}
\end{table}

Their hidden and feedforward dimensions grow with scale, while the head dimension stays fixed and the number of key-value heads is held at eight across all three.
This lets us separate the effect of the hardware from that of the architecture, and see how a single design behaves as it moves across generations of a device.

\subsection{Runtime Analysis}
\label{ssec:qwen-other}

\begin{figure}[h!]
	\centering
	\includegraphics[width=\linewidth]{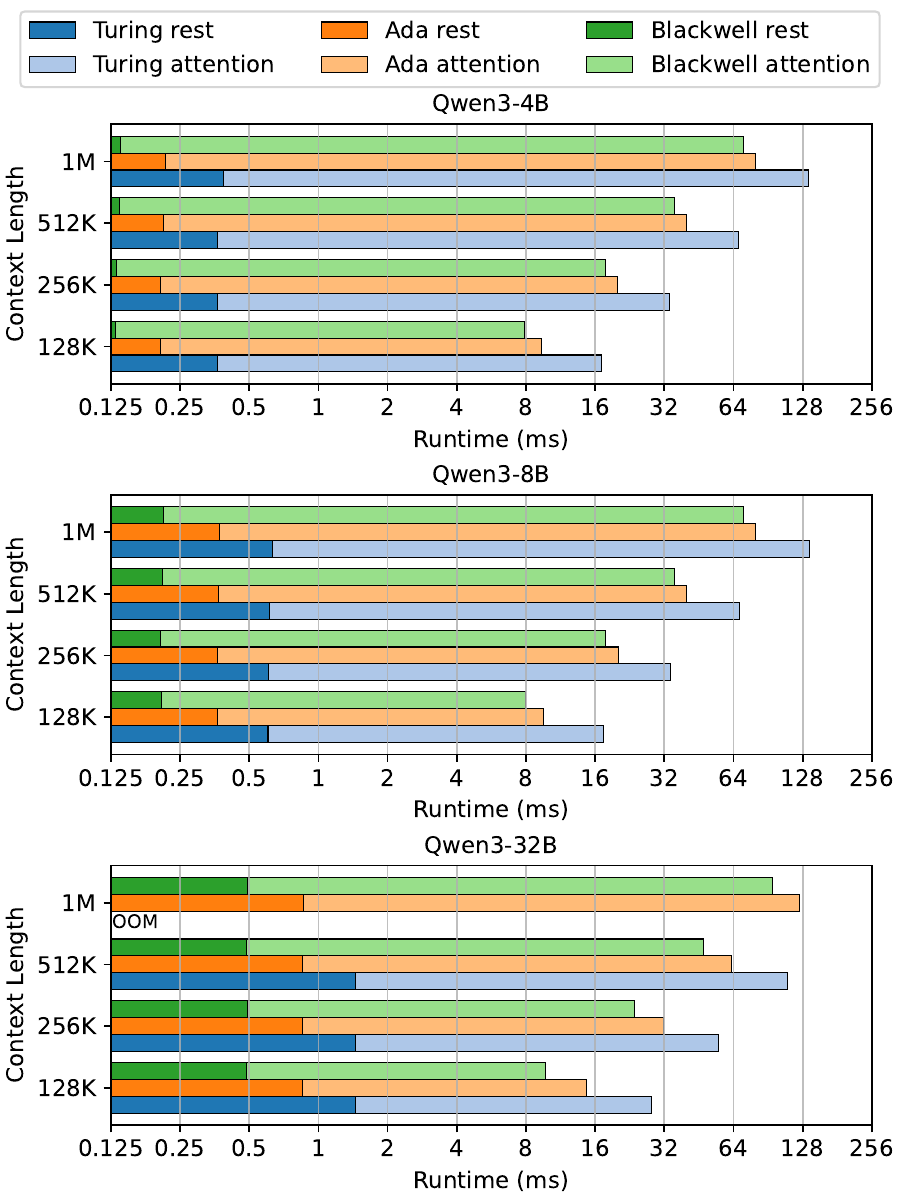}
	\caption{Runtime of self-attention vs. the rest of the block for the Qwen3 LLMs on each GPU.}
	\label{fig:qwen3:runtime-analysis}
\end{figure}

The runtime of a block is made of two parts that behave differently, and each new GPU acts on them unequally.
\Cref{fig:qwen3:runtime-analysis} reports both for the three models on each of the three GPUs, as the context grows.
As before, we extend the context beyond the length the models support, which remains faithful for the reasons given in \Cref{ssec:gpt:rooflines}.

We consider first the rest of the block, shown as the darker segment of each bar.
This part does not grow with the context, so it isolates what each new GPU contributes to runtime.
Across the three devices, each generation is faster than the last, with Ada around 1.7$\times$ faster than Turing and Blackwell between 1.5$\times$ and 1.75$\times$ faster than Ada, a gain of up to 3$\times$ from the oldest device to the newest.
These gains, however, apply to the part of the runtime that stays fixed.
As the context grows, the attention segment grows with it, while this part does not.

We turn now to the attention segment, the lighter part of each bar.
As the context grows, it comes to occupy most of the runtime, and by one million tokens it accounts for nearly the entire bar across all three models, a length at which Qwen3-32B no longer fits on the Turing GPU.
The gains of each new device fall on the rest of the block, which by this point contributes little, so a faster GPU brings little benefit where the runtime is actually spent.

\begin{tcolorbox}
    \textbf{Observation \#5:} A block's runtime splits into a part that grows with the context, the attention computation, and a part that stays fixed, the remaining kernels. Newer GPUs accelerate only the fixed part, so their advantage is confined to short contexts.
\end{tcolorbox}

\subsection{Compute and Bandwidth Utilization}

\begin{figure}[h]
	\centering
	\includegraphics[width=\linewidth]{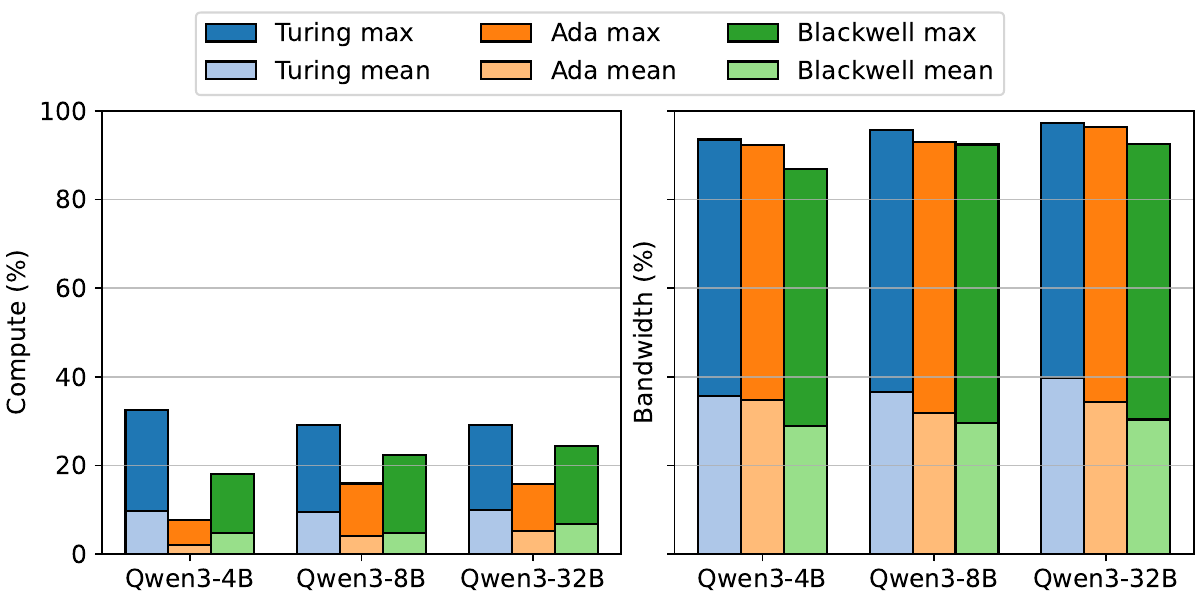}
	\caption{Mean and maximum compute and bandwidth utilization across Qwen3 models and GPUs for $L=131,072$.}
	\label{fig:qwen3:compute-bandwidth}
    \vspace{1em}
\end{figure}

The question we now ask is whether model sizes impact the resource utilization of GPUs.
Likewise, we examine whether these devices are efficiently utilized and if there are differences between them.
Figure~\ref{fig:qwen3:compute-bandwidth} displays the amount of compute and bandwidth each device uses for the three Qwen variants.
Measurements are taken at $L=131,072$, the maximum context length supported by the models.

Bandwidth is well utilized overall.
Each device achieves a maximum close to the theoretical throughput across the three Qwen3 sizes.
We observe a decrease of a few percentage points with each new generation of GPU.
The mean bandwidth used is just below 40\%, following the same downward trend with each new device.

Meanwhile, the results obtained for compute are fairly low.
The oldest GPUs use the most compute, with maxima around 30\%.
Ada obtains the lowest compute utilization, with a mean below 5\% and a maximum below 15\%.
In comparison, the latest microarchitecture obtains a slight increase.

When considering both resources, model size does not increase device utilization.
The oldest GPU uses more of the available hardware resources compared with the other two.
However, bandwidth and compute usage are unbalanced across GPU generations.
Local inference optimizations should target this imbalance, leveraging the unused compute to improve performance.

\begin{tcolorbox}
	\textbf{Observation \#6:}
	There is a major imbalance in the utilization of GPU resources.
    Newer GPUs with faster bandwidth and variations in mode size do not address those problems.
\end{tcolorbox}

\subsection{Attention Across Generations}
\label{ssec:quen-attentions}

In the previous section, each generation of hardware delivered a gain, and for the fixed part of the block, that gain is roughly three times from the oldest device to the newest.
That part, however, is not where the runtime is spent in long contexts.
Asking the same question of attention gives a different answer.
Between Turing and the two newer devices, attention improves as expected.
As \Cref{fig:qwen3:runtime-analysis} shows, the improvement between Ada and Blackwell is around 12\% for Qwen3-4B and Qwen3-8B at the longest contexts, and 32\% for Qwen3-32B.

The three models differ in one respect.
Qwen3-32B has 64 query heads, whereas the other two have 32, and the number of key-value heads is eight in all three.
Blackwell pulls ahead of Ada on the model that gives the kernel twice as many heads to work with, and not on the two that do not.

Blackwell has more bandwidth, more cores, and a larger cache than Ada.
The fixed part of the block turns each of these into speed, while attention does not.
\Cref{fig:qwen3:compute-bandwidth} shows that Blackwell draws more compute than Ada while running attention no faster.
The kernel is therefore not failing to use the newest device for want of resources.
It is using them without the gain that the previous generation obtained from less.

\begin{tcolorbox}
	\textbf{Observation \#7:}
	Our results indicate that the attention kernel is not as optimized for the newest microarchitecture.
	The rest of the block has a faster runtime with each new GPU for small and large models.
	However, for the smaller LLMs and when the context is long, the Blackwell and Ada GPUs have a runtime that is almost the same.
\end{tcolorbox}

\section{Recommendations}
\label{sec:recommendations}

The observations of the previous two sections describe a device that is rarely used well.
Attention dominates the runtime and yet reaches neither of the device's limits.
The kernels around it are worse, running briefly and using few resources, and there are ever more of them as architectures grow more elaborate.
Newer hardware does little to change this, since what it accelerates is precisely the part that matters least.
If the device is to be used better, the change has to come from the software that drives it.
We recommend four such changes below.

\subsection{Improving Attention Kernels for Long Contexts}
\label{ssec:mainstream-long-attention}

\begin{figure}[h]
	\centering
	\begin{subfigure}[b]{\linewidth}
		\centering
		\includegraphics[width=\textwidth]{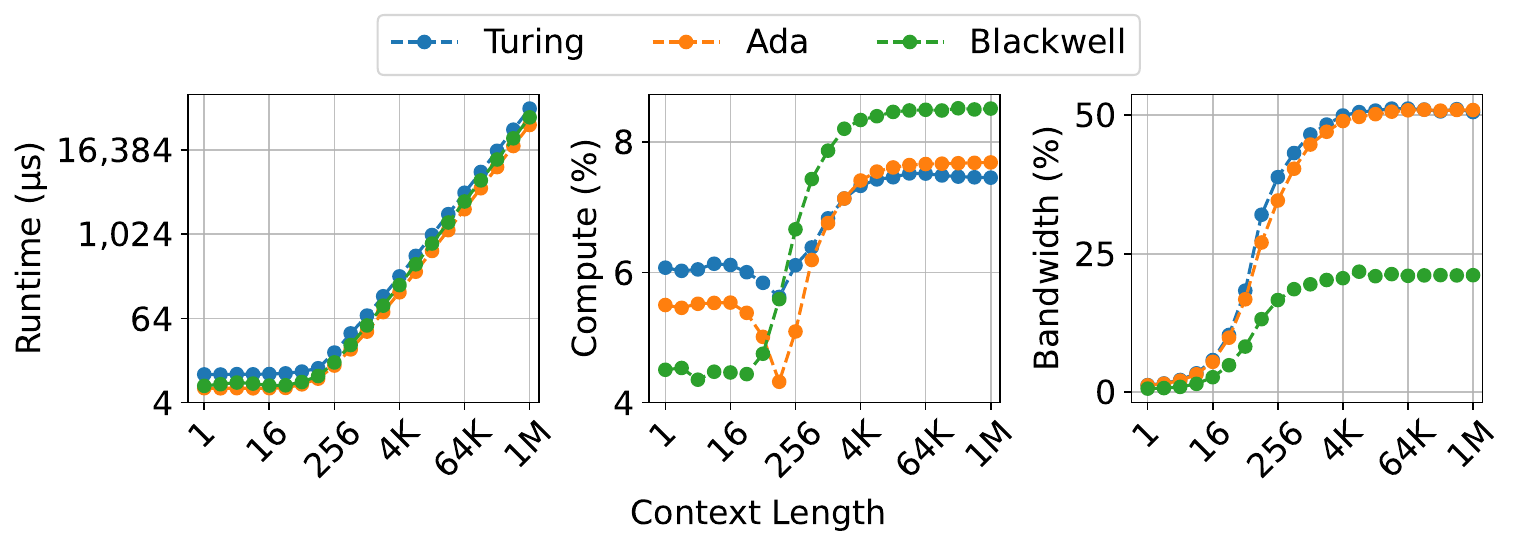}
		\caption{Qwen3-8B}
		\label{fig:recommendations:attention:qwen3-8b}
	\end{subfigure}
	\medskip
	\begin{subfigure}[b]{\linewidth}
		\centering
		\includegraphics[width=\textwidth]{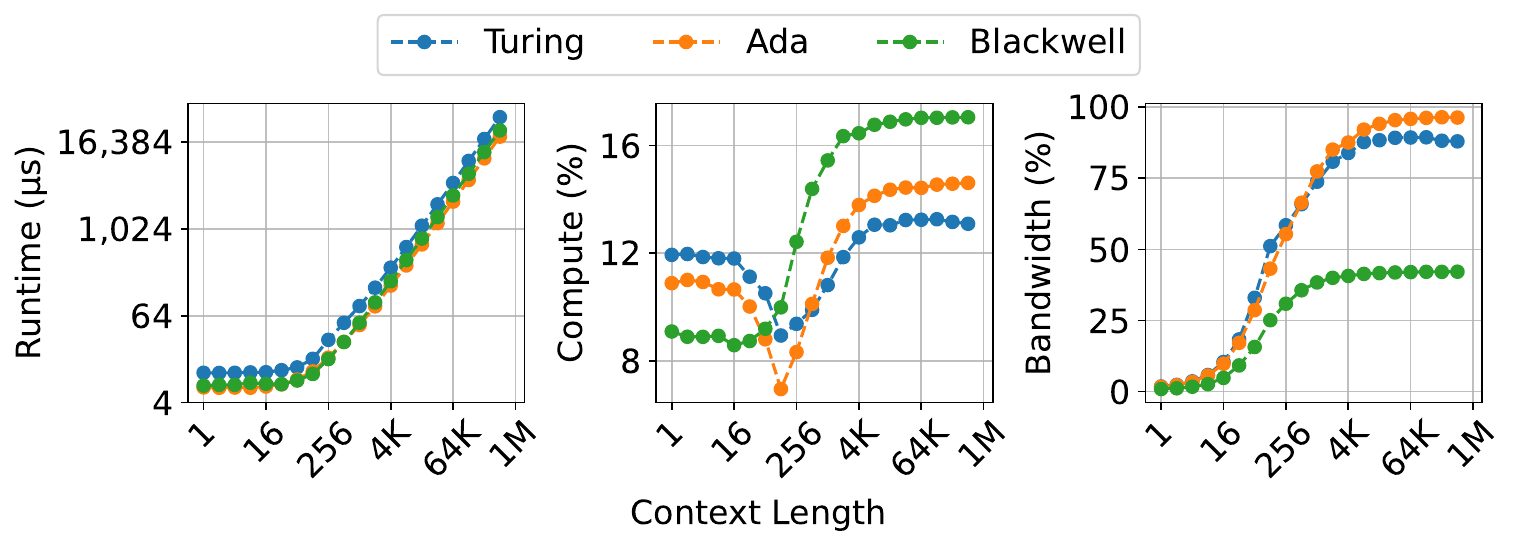}
		\caption{Qwen3-32B}
		\label{fig:recommendations:attention:qwen3-32b}
	\end{subfigure}
	\caption{Runtime, bandwidth, and compute for Qwen3's attention kernel over increasing context length.}
	\label{fig:recommendations:attention}
\end{figure}

Attention is the operation that grows with the context, so it is the one whose use of the device matters most at long contexts.
\Cref{fig:recommendations:attention} reports its runtime, compute, and bandwidth for Qwen3-8B and Qwen3-32B across the full range of context lengths, on each of the three GPUs.

As the figure shows, the runtime rises as expected, but neither resource reaches the devices' limits.
Compute stays below 9\% for Qwen3-8B and below 17\% for Qwen3-32B.
In turn, the bandwidth reached at long contexts differs sharply across the devices, from near saturation on Turing and Ada to a mere 21\% on Blackwell for the smaller model.
The underutilization is not caused by the volume of data moved during decoding. 
As the context grows, the KV cache dominates what the kernel reads, and yet the bandwidth to move it goes unused. 
The real issue is that the kernel does not expose sufficient parallelism to fully utilize the GPU.

The required parallelism for the kernel must be sourced from some dimension.
With a batch size of one and a fixed number of heads, it is the sequence that provides this opportunity for parallelism.
For example, Flash-Decoding~\cite{daoflashdecoding} introduces an additional dimension of parallelism by dividing the KV cache into segments referred to as splits. 

With this partitioning, Attention is computed in parallel between the query and each split, and a final reduction step combines the results.
Splitting the KV cache in this way has the potential to exploit more of the parallelism available on modern GPUs. 
However, both this approach and other parallelization techniques (\eg parallelizing over the query rather than the KV cache) have not, to our knowledge, been studied systematically.

Beyond the lack of study, these techniques are also unevenly available.
Flash-Decoding exists for NVIDIA GPUs, and is relatively newer for AMD, where the implementations carry heuristics tuned for NVIDIA hardware.
On CPU targets, it does not exist at all.
Moreover, even where the technique is available, the compiler framework does not choose it automatically.
The attention kernel is chosen based on a predetermined order of preference rather than the specific characteristics of the problem, such as the presence of a single query. 
Users who wish to enable KV cache splitting must first be aware of its existence and then request it explicitly.

Navigating these options should be done by the compiler, not by the user.
The latter should use the fact that the kernels in question compute the same attention, but differ in how they distribute the work across the device.
The choice therefore depends on the number of users in the workload, and the compiler should be able to detect that and choose the right kernel.
Currently, it does not.

\begin{figure*}[t]
    \centering
    \includegraphics[width=0.9\linewidth]{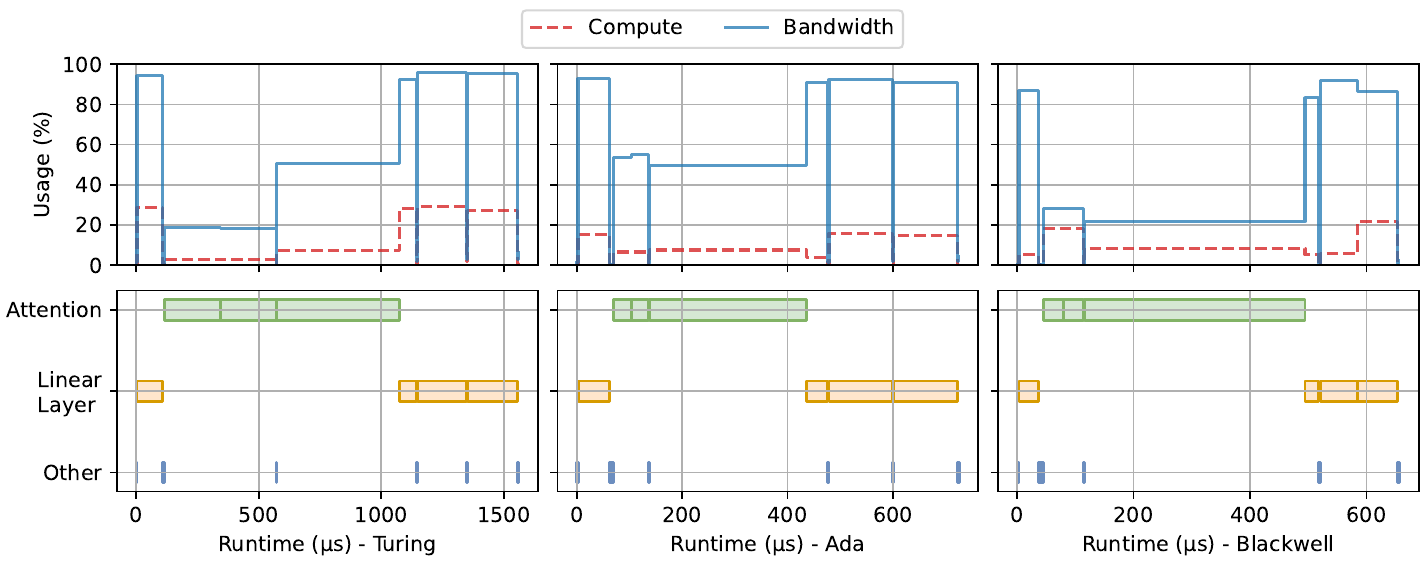}
    \caption{Timeline view of Qwen3-8B ($L=8,192$).}
    \label{fig:recommendations:qwen3-timeline-view}
\end{figure*}

\begin{tcolorbox}
    \textbf{Recommendation \#1:} Parallelizing attention along the sequence axis should become a mainstream technique for long contexts.
    Variations of this approach warrant systematic study, implementations should support a broader range of hardware, and the compilation stack should select this method based on the problem’s shape rather than leaving the choice to the user.
\end{tcolorbox}

\subsection{Precise Paging Opportunities}
\label{ssec:paging}

Our observations so far were limited to how the device is used, not whether the data fits on it.
The single block we profile makes capacity a non-issue by construction, and a full model at a long context does not enjoy this luxury.
One answer to that problem is paging, and its benefit has been established since vLLM introduced block-based memory management~\cite{kwon2023efficient}.
The purpose there, however, is not the one we have in mind.
In a system serving many users at once, paging resolves contention, evicting the blocks of sequences that are not currently running so that others may proceed.
With a single query, there is no victim worth evicting, since every block it holds is one it needs for its next step.
Paging has also been used to a second end, fitting a model larger than the device onto a single GPU, and FlexGen is the best known of these~\cite{sheng2023flexgen}.
It is this second use that our proposal resembles.

What matters for such a use is the granularity at which paging occurs.
FlexGen operates at the granularity of a layer, and recognizes that the data of the next layer must be prefetched while the current one computes.
FlexGen is built for many queries at once and pursues throughput, accepting a cost in latency to obtain it.
We ask instead what can be done for a single user, where latency is the quantity that matters and cannot be traded away.

\begin{figure}[t]
	\centering
	\begin{subfigure}[b]{\linewidth}
		\centering
		\includegraphics[width=\textwidth]{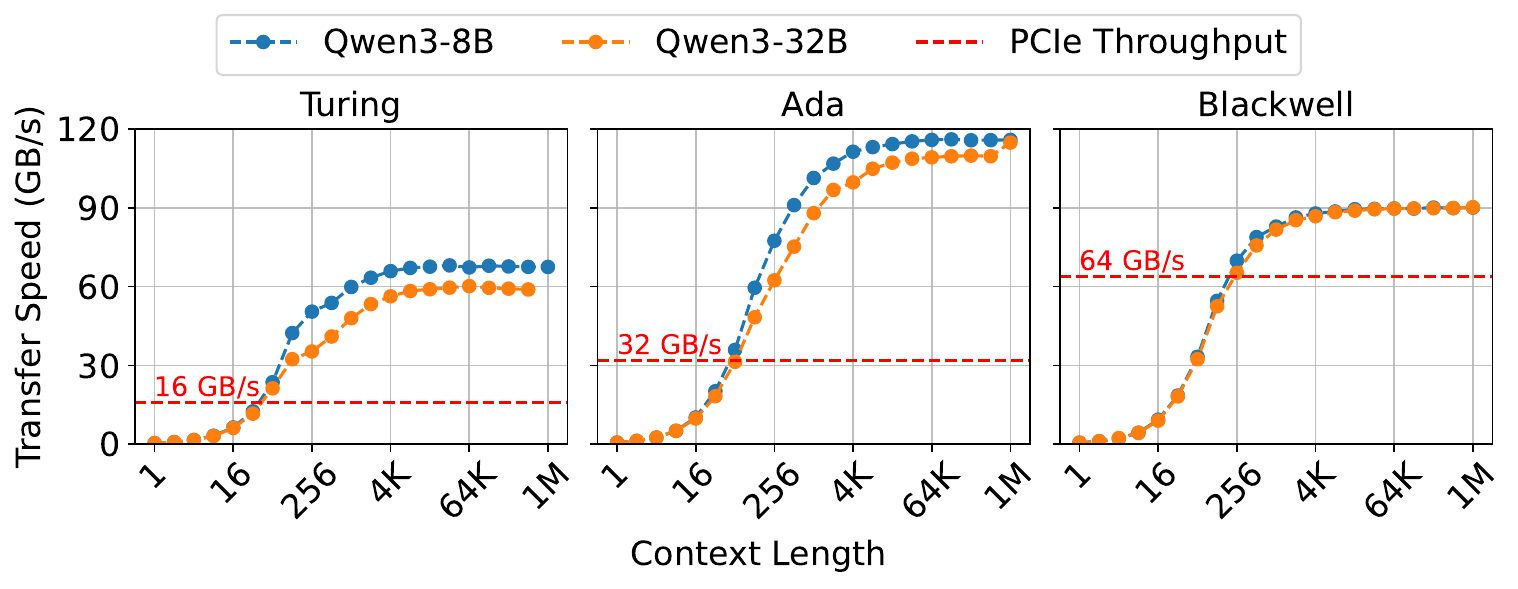}
		\caption{Attention only}
		\label{fig:discussion:transfer-speed:attn}
	\end{subfigure}

	\medskip
	\begin{subfigure}[b]{\linewidth}
		\centering
		\includegraphics[width=\textwidth]{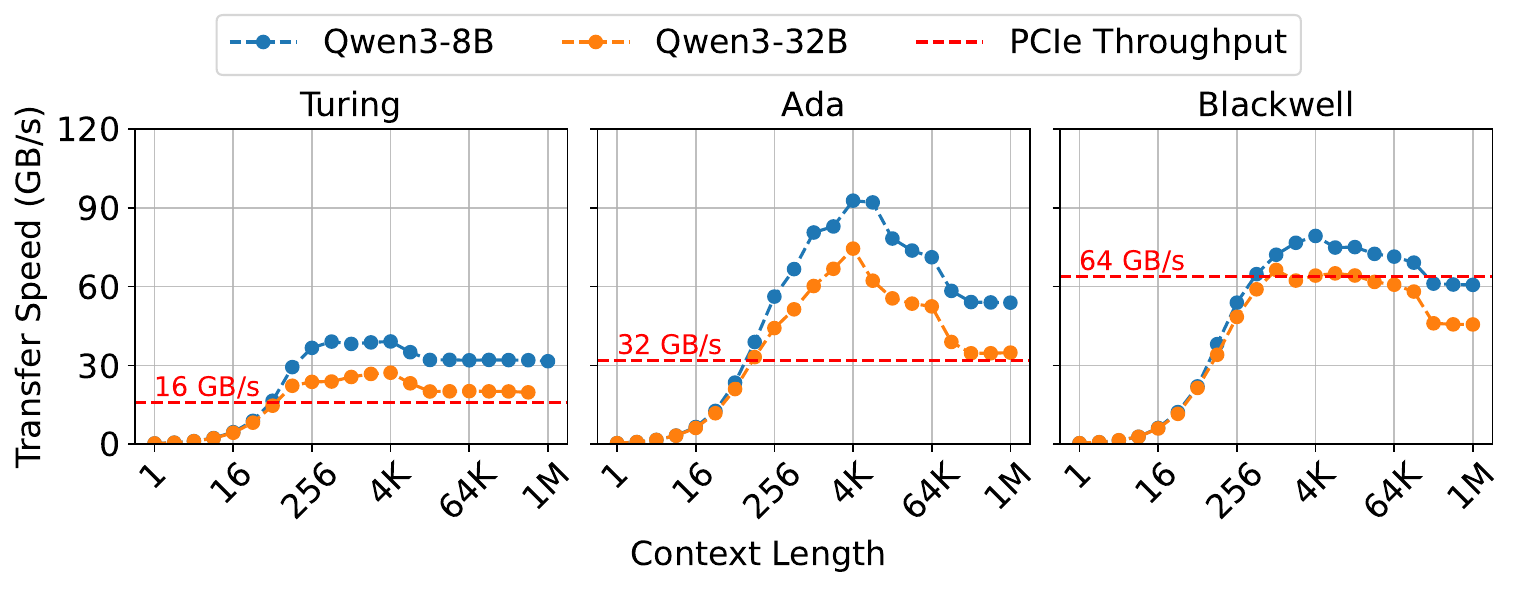}
		\caption{Attention \& repeat kernels}
		\label{fig:discussion:transfer-speed:attn-repeat}
	\end{subfigure}
	\caption{Transfer speed required to page the KV cache of the Qwen3 LLMs.}
	\label{fig:discussion:transfer-speed}
\end{figure}

Answering that question requires knowing when a transfer may take place without competing for memory bandwidth with the kernels themselves.
\Cref{fig:recommendations:qwen3-timeline-view} shows where that moment lies.
Once the GPU has loaded the KV cache and turns to computing attention, its use of memory bandwidth falls, and it stays low for the duration of the computation.
This is a window during which data can move without contending with anything.
The proposal is therefore not to schedule transfers on demand, but to trigger them at a point where we know in advance they will not interfere.

The window is of use only if it is long enough to page in the data of the next layer.
\Cref{fig:discussion:transfer-speed} reports the transfer speed that would be required to do so, against what the PCIe interface of each device provides.
Taking the attention computation alone, shown in \Cref{fig:discussion:transfer-speed:attn}, the required speed exceeds what PCIe supports.

However, when we consider the window that includes the kernels that prepare attention, the copies that grouped-query attention requires, the picture changes.
As \Cref{fig:discussion:transfer-speed:attn-repeat} shows, the required speed then falls close to what PCIe provides on Ada and below it on Blackwell.
The window is sufficient, and it is on the newer platforms that it is most comfortably so.
The implication is that there can be a form of paging that costs little to no latency, which would remove the memory of a single device as the limit on context length.
Realizing it with current hardware is another matter.
This opportunity could be seized by using storage access from the GPU through GPUDirect~\cite{gpudirectstorage}.

The issue, however, is that such technology does not allow it to be triggered from the precise point inside the computation where the window opens.
What the mechanism calls for is a memory or storage device reachable over peer-to-peer PCIe, able to serve the GPU directly and at a moment the GPU itself determines.

\begin{tcolorbox}
    \textbf{Recommendation \#2:} Paging the KV cache out of GPU memory can be made to interfere little with the computation by triggering the transfer where the device leaves bandwidth idle, during attention and the kernels that prepare it.
    Hardware should allow the GPU to initiate such transfers at the moment it determines.
\end{tcolorbox}

\subsection{Improving Compiler Fusion}

ML compilers offer several advantages for inference over \textit{eager mode} frameworks~\cite{tokui2019chainer, paszke2019pytorch}.
One major optimization performed by the compiler is operation fusion, such as combining the up projection and the activation function of the FFN into a single kernel.
The autotuner can evaluate different implementations, such as Triton kernels~\cite{tillet2019triton} and those from libraries like NVIDIA cuDNN or cuBLAS~\cite{chetlur2014cudnn, nvidiacublas}.
Likewise, it can benchmark different tile sizes for loading data from global memory and different thread block configurations.

In our studies, we find that the compiler performs almost no operation fusion across GPU generations and models, resulting in an optimized version that still closely mirrors the mathematical description.
This problem is exacerbated by the architectural evolution of LLMs: newer and more performant layers introduce additional operations.
For example, MoE introduces a gate layer and a top-$k$ operator to select the experts, two small steps that are not combined.
This lack of fusion results in numerous kernels and introduces two main issues.
First, each kernel launch has some launch overhead~\cite{hijma2023optimization}.
Second, implicit synchronization is injected between each of the kernels, forcing the execution of a kernel to complete before starting the next one.
As a result, SMs are not continuously and proactively loading data from memory, underutilizing a memory bandwidth that remains the main bottleneck of inference speed.

\begin{figure}[t]
    \centering
    \includegraphics[width=\linewidth]{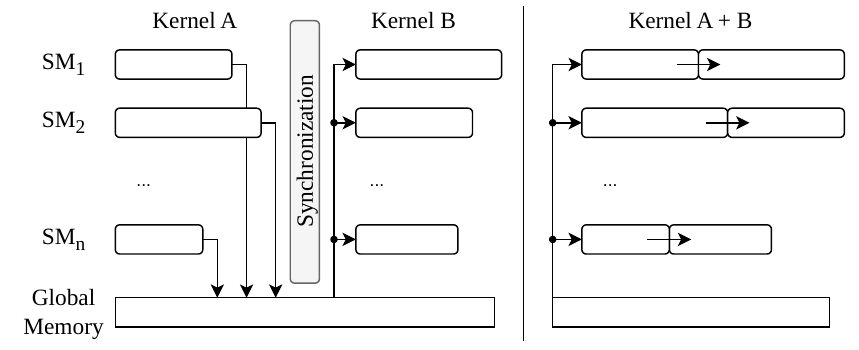}
    \caption{Implicit synchronization and materialization of two kernels (left) vs. the fused version (right).}
    \label{fig:discussion:fusion}
\end{figure}

ML compilers should be able to fuse a larger number of operators together, vastly reducing the number of kernel launches.
This would resemble the megakernel approach to fusion~\cite{nrusimha2025flashformer, spector2025look}, combining a full model into a single kernel.
This allows threads to proceed to the next step of the computation, instead of having to wait at each layer, illustrated in Figure~\ref{fig:discussion:fusion}.
Current approaches are handwritten and require explicit synchronization carefully placed by kernel experts.
Work has been done to build compilers capable of generating such kernels, such as MPK~\cite{cheng2026mpk}.

\begin{tcolorbox}
	\textbf{Recommendation \#3:}
	The number of kernels has increased in modern LLM architecture, and many are not fused.
	ML compilers should fuse more operations together to reduce synchronizations and improve device utilization.
\end{tcolorbox}

\subsection{Reducing Kernels Stalls}

Self-attention is the most costly operation during the decode phase on long contexts.
As displayed in Figures~\ref{fig:gpt-models:attention-rooflines} and~\ref{fig:recommendations:attention}, our studies indicate that this kernel does not use all the GPU resources available and, in particular, only a fraction of the available compute.
To better understand part of the reason behind these results, we analyze the assembly language of NVIDIA GPUs, SASS, to find the causes of stalls.
A cycle is stalled when no instruction can be issued by the SM, forcing threads to wait~\cite{alsop2016gsi}.
Table~\ref{tab:discussion:stalls} shows the ten instructions causing the most stalls in the attention kernel of GPT-OSS when compiled for the Blackwell 6000 GPU.

\begin{table}[t]
	\caption{Top 10 instructions causing stalls and their main reasons for the attention kernel of GPT-OSS ($L=131,072$) on the Blackwell GPU.}
	\centering
	\label{tab:discussion:stalls}
    \resizebox{\columnwidth}{!}{%
	\begin{tabular}{lcl}
		\toprule
		SASS Instruction & Stalls Count & Main Reason \\
		\midrule
		\texttt{BAR.SYNC.DEFER\_BLOCKING 0x0} & 33,689 & Long Scoreboard \\
		\texttt{@!P3  LDS.U16 R84, [R84+0x600]} & 23,306 & Barrier \\
		\texttt{STS.128 [R91+0x600], R16} & 7,197 & Long Scoreboard \\
		  \texttt{LDSM.16.M88.4 R68, [R0+UR5+0x600]} & 2,690 & Barrier \\
		\texttt{HMMA.1688.F32.BF16 R36, R4, R8, R36} & 1,614 & Short Scoreboard \\
		\texttt{@!P3  IMAD.U32 R5, R84, 0x10000, RZ} & 1,440 & Short Scoreboard \\
		  \texttt{BSSY.RECONVERGENT B0, 0x70113372f8f0} & 1,248 & Barrier \\
		\texttt{HMMA.1688.F32.BF16 R4, R68, R72, R4} & 1,180 & Short Scoreboard \\
		\texttt{ULEA UR4, UR4, UR25, 0x18} & 1,090 & Short Scoreboard \\
		\texttt{FSETP.GEU.AND P0, PT, R109, R6, PT} & 1,067 & Short Scoreboard \\
		\bottomrule
	\end{tabular}%
    }
    \vspace{-1em}
\end{table}

\begin{figure}[h]
    \centering
    \includegraphics[width=0.9\linewidth]{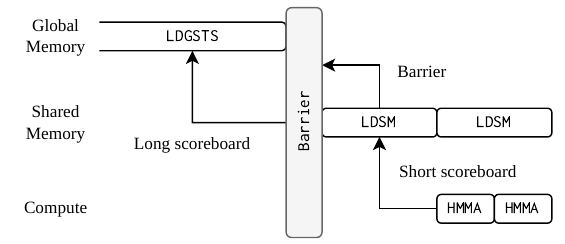}
    \caption{Simplified view of stalls in GPU execution.}
    \label{fig:discussion:stalls}
\end{figure}

For the attention operation, long and short scoreboards as well as barriers account for the majority of stalls.
Long scoreboards are delays caused by loads from global memory, whereas short scoreboards are caused by loads from shared memory.
Barriers are a synchronization primitive that blocks the execution of threads, forcing them to wait for each other.
Furthermore, long stalls are often caused by multiple ones beforehand: their scoreboard dependencies often include barriers.
For example, when looking at the dependency on the fifth instruction with the most stalls, a half-precision matrix multiply and accumulate (\texttt{MMA}), we find that it depends on a load from shared memory (\texttt{LDSM}).
When investigating what caused this load to stall, we found that this instruction depends on a barrier synchronization (\texttt{BAR.SYNC.DEFER\_BLOCKING}).
Figure~\ref{fig:discussion:stalls} shows how stalls can accumulate, halting execution of the GPU.
Here, the MMA instruction cannot be executed, as it must first wait for a load from shared memory (\texttt{LDSM}).
This instruction cannot be initialized yet since it is waiting on a barrier, necessary to synchronize threads and ensure data loaded from global memory (\texttt{LDGSTS}) has arrived in shared memory.

Kernels should be designed in a more asynchronous manner such that threads continuously perform work instead of waiting for each other or for data to arrive.
Addressing kernel stalls could help increase compute utilization.
Tensor cores are not useful to address this problem, as they do not help improve the performance of memory-bound workloads~\cite{zhang2025can}.
Part of the solution comes from a reduced use of synchronization and more fine-grained timing of memory accesses.
The new hardware unit introduced with the Hopper microarchitecture and available in Blackwell can help with data movement.
The Tensor Memory Accelerator (TMA unit) can efficiently transfer data from global memory to shared memory asynchronously~\cite{patel2025nvidia}.
None of the generated kernels running on the RTX PRO 6000 Blackwell GPU made use of this hardware unit.

\begin{tcolorbox}
	\textbf{Recommendation \#4:}
	Threads are forced to wait on each other because of synchronization, slowing the computation.
	Kernels should be designed around this problem to reduce the number of stalls.
\end{tcolorbox}

\section{Discussion}
\label{sec:discussion}

Our measurements are now in hand, and so are the recommendations we draw from them.
The recommendations, however, raise questions that the numbers alone do not answer.
In the following sections, we present three of them: whether the problems behind them are ones this community has met before, whether compression remains necessary once the device is used well, and what kind of system would be needed to act on any of it.

\subsection{Recurring Problems in a New Setting}
\label{ssec:recurring-problems}

An LLM serving system is not a database, and nothing in the recommendations above asks that it be treated as one.
However, we claim a similarity in the issues those two systems encounter.
We now substantiate this claim by relating each of our recommendations to prior work from this community.

Let us start with Recommendation \#2.
The proposal that the interconnect be treated as a resource the system schedules rather than a channel it merely uses has been advanced in this community, resting on the observation that a workload starved of one resource leaves another idle~\cite{jiang2026data}.
Our paging window applies that principle within a single computation rather than across competing ones.

Recommendations \#3 and \#4 find their counterpart in a characterization of four GPU database systems~\cite{cao2023gpu}.
That study attributes the underutilization it measures to a lack of kernel fusion, the cause behind our third recommendation.
The remedy proposed alongside it, a set of routines that compose without materializing between them~\cite{shanbhag2020study}, is what we ask of a compiler when we ask it for fewer and larger kernels.
The same study also finds schedulers left idle by memory waits, which account for part of the stalls we found.

Recommendation \#1 is the exception, in that the case for parallelizing attention along the sequence rests on our measurements alone.
The portability it also calls for, however, does not.
The argument that an accelerated engine should not be written against the hardware of one vendor has been made here at production scale, with hardware independence adopted as a design principle and validated across accelerators and interconnects of several kinds~\cite{interlandi2026coddspeed}.

These difficulties were met before on a different workload, query processing.
We believe that the experience gained in those contexts can now be applied, from a systems perspective, to models that generate one token at a time for a single user.

\subsection{On the Necessity of Compression}
\label{ssec:no-compression}

Compression exists to make a model fit on the device.
Much of the work in this direction reports on the footprint achieved, with weight compression now approaching one bit per parameter.
Less is said about what is given up in return, though the evidence that something is given up is not scarce: under tasks that demand the knowledge a model holds, pruning degrades performance at ratios as low as 25 to 30\%, and quantization to eight bits already costs eight to ten percent~\cite{jaiswal2024compressing}.
That loss grows sharper as the context a model must digest grows longer, which is the regime this paper concerns itself with.
Recommendation \#2 offers a different way to make a model fit, one that pages the dominant and momentarily unused part of the state in and out of the device rather than shrinking it, and pays for the transfer with bandwidth that would otherwise sit idle.
Where that window suffices, compression may indeed prove not to be necessary or at least as important.

That window, however, is not fixed.
It exists because attention is slow, and Recommendation \#1 asks that attention be made faster.
Were the other recommendations carried out, the time available to page would shrink along with the runtime they recover, and at some point paging would cost latency rather than come free.
We do not claim, then, that a workstation will serve a long context as fast as the datacenter hardware built for the task.
Our claim is that the difference may matter less than it appears, because a single user running an agentic workload is rarely waiting on the next token.
Such workflows proceed asynchronously, and a model that generates somewhat more slowly, on a machine the user owns, with data that never leaves it, may still be a worthy tradeoff.

\subsection{The Case for a New Single-User Serving System}
\label{ssec:new-system}

A decoder block is not a serving system.
One may argue, then, that what we measured on a block says little about what a system would do.
We would answer that a system has no way of avoiding what the block does.
Since LLMs are just a repetition of blocks, improving the latter is how we improve the former.
We believe that our recommendations can be such a way to improve local serving systems.
Moreover, we believe the recommendations are fundamental enough to justify building a system around them, as opposed to retrofitting them in existing systems.

This raises the question of why not PyTorch.
The answer is not that PyTorch is at fault.
A compiler emits general code because it cannot know which model, which device, and which regime it is being asked for, and that knowledge lives in the layer above it.
vLLM shows what the layer above is worth: it compiles through \texttt{torch.compile} as we do, and by telling the compiler what it is compiling for, obtains kernels that a general invocation does not.

Naturally, why not vLLM, then?
What it tells the compiler is that many queries will arrive at once, and that premise runs through the system rather than sitting at its edge.
Paging~\cite{kwon2023efficient}, continuous batching~\cite{yu2022orca}, and prefix caching~\cite{zheng2024sglang} all resolve contention among those queries, and the shapes it specializes for systems serving multiple requests.
A single-user scenario lacks these characteristics, and expecting a system to perform efficiently outside its intended design parameters does not make effective use of its strengths.

The lineage that does take the single user as its premise begins with llama.cpp~\cite{gerganov2023llamacpp}, and it has carried local inference a long way.
What has not followed from it is a system as production-ready for one user as vLLM is for many.
It may be time for one, and our recommendations describe the pillars on which the architecture of this new potential system should be based.
We hope to see such a system because, as is, our findings indicate that GPU architecture improvements are not yet directly benefiting the single-user inference workload.

\section{Related Work}
\label{sec:related}

Studies have looked into characterizing LLM inference workloads.
Recasens et al. analyze small language model inference with the goal of finding the optimal batch size to maximize tokens per second and device utilization~\cite{recasens2024pareto}.
Inference has been found to remain memory-bound even for large batches, and never maximizes compute utilization~\cite{recasens2025mind}.
Pope et al. study the inference latency of transformer models with more than 500B parameters, and how to scale it to multiple TPUs~\cite{pope2023efficiently}.
Inference of some Llama models has been studied through the roofline model, to determine if the different layers are memory-bound or compute-bound~\cite{yuan2024llm}.
The workloads of encoder and decoder transformer models have been studied for both training~\cite{gholami2024ai} and inference~\cite{kim2023full}.
It remains, though, that these studies have not focused on single-user workloads, as ours does, and in particular in the face of increasing context lengths.

Other work specifically focuses on kernels and ML compilers.
 Carmona-Martínez et al. compare the productivity, portability and performance of ML compilers for inference on consumer GPUs~\cite{martinez2026characterization}.
Attention kernels have also been compared during training for long contexts and distributed settings~\cite{bu2025longcontext}.
These studies are complementary to ours and may indicate how some of the recommendations, such as kernel fusion, may be better achieved by newer techniques or by compiler improvements.

Lastly, studies have looked at how the energy efficiency of transformer models and LLMs~\cite{wang2020energy, fernandez2025energy}.
The energy efficiency of deep neural networks has also been studied for neural network training~\cite{you2023zeus}.
We believe that those studies reflect some of the poor resources utilization that our study finds as well.
Improving the utilization of resources on a GPU should, in theory, improve the value gained for each joule invested.

\section{Conclusion}
\label{sec:conclusion}

In this work, we set out to investigate the obstacles that prevent a long-context LLM from being served locally on a single GPU for individual users.
The prevalent assumption is that such a model must be heavily compressed, or else left to a provider.
We expected, then, to find a GPU device overwhelmed by the workload.
Instead, we observed the opposite.
If we set the capacity issue aside, the device had resources to spare, and this was not the measure of one model on one GPU.
Across two families of models and three generations of hardware, no device was ever pushed to its limits.

A natural next question was, then, how to make better use of the available resources.
The opportunities we identify are varied, and no single solution solves all the inefficiencies found.
Solving the capacity issue, however, is one of these opportunities.
A particular window of memory bandwidth availability opens during attention.
It stays open long enough to page the cache off the device, allowing for longer context lengths to be used.
Compression thus appears to be a concession to how the device is currently driven rather than a requirement of the workload.
More generally, we found some opportunities for improvement that may inform a new generation of single-user serving systems to be built.
In particular, many of the problems involved in building such a system have already been encountered by this community in GPU query processing, and the experience gained there may prove influential here as well.

\balance

\bibliographystyle{ACM-Reference-Format}
\bibliography{main-sanitized}

\end{document}